\documentclass[11pt]{article}
\usepackage{amssymb,epsfig,theorem}
\newcommand{\nc}{\newcommand}
\nc{\be}{\begin{equation}}
\nc{\ee}{\end{equation}}
\nc{\bea}{\begin{eqnarray}}
\nc{\eea}{\end{eqnarray}}
\nc{\disp}{\displaystyle}
\nc{\ade}{\mbox{$A$-$D$-$E$}}
\nc{\calN}{{\cal N}}
\nc{\calC}{{\cal C}}
\nc{\calM}{{\cal M}}
\nc{\calS}{{\cal S}}
\nc{\phit}{\hat{\varphi}}
\nc{\chit}{\hat{\chi}}
\nc{\hcalN}{\hat{\calN}}
\nc{\hcalS}{\hat{\calS}}
\nc{\hS}{\hat{S}}
\nc{\sigmad}{\sigma^\dagger}
\nc{\psid}{\psi^\dagger}

\def\sstyle{\scriptstyle}
\def\dps{\displaystyle}
\def\sh{{\rm sh}}
\def\ch{{\rm ch}}
\def\cth{{\rm cth}}
\def\th{{\rm th}}

\font\tenmsb=msbm10\font\sevenmsb=msbm7
\font\fivemsb=msbm5
\newfam\msbfam
\textfont\msbfam=\tenmsb
\scriptfont\msbfam=\sevenmsb
\scriptscriptfont\msbfam=\fivemsb
\def\dps{\displaystyle}

\def\bra#1{\langle #1|}
\def\ket#1{|#1\rangle}

\def\i{{\rm i}}

\newcommand{\sectiona}{\setcounter{equation}{0}\section}

\begin{document}
%\draft
%\preprint{}
\title{Bethe Ansatz for the Temperley-Lieb loop model with open boundaries}

\author{Jan de Gier$^1$ and Pavel Pyatov$^2$.
\\[5mm]
{\small \it $^1$Department of Mathematics and Statistics, University
of Melbourne, Parkville,}\\  {\small \it Victoria VIC 3010, Australia.} \\ 
{\small \it$^2$Bogoliubov Laboratory of Theoretical Physics, Joint
  Institute for Nuclear Research,} \\ {\small \it Dubna, Moscow Region
141980, Russia.}
}

\date{\today}
\maketitle

\begin{abstract}
We diagonalise the Hamiltonian of the Temperley-Lieb loop model with open
boundaries using a coordinate Bethe Ansatz calculation. We find that in the
groundstate sector of the loop Hamiltonian, but not in other sectors,
a certain constraint on the parameters has to be satisfied. This
constraint has a natural interpretation in the Temperley-Lieb algebra
with boundary generators. The spectrum of the loop model contains that
of the quantum spin-$1/2$ XXZ chain with non-diagonal boundary
conditions. We hence derive a recently conjectured solution of the 
complete spectrum of this XXZ chain. We furthermore point out a
connection with recent results for the two-boundary sine-Gordon model.  
\end{abstract}

\sectiona{Introduction}
\label{se:intro}
It has been known for some time that the quantum spin-$1/2$ XXZ chain with
non-diagonal boundary fields is integrable \cite{Sklya88}. However,
conventional methods to calculate its spectrum, such as the Bethe
Ansatz, have been found difficult to apply due to a lack of spin
conservation or other good quantum numbers. Recently however,
Nepomechie made progress using functional equations. In root of unity
cases he obtained Bethe Ansatz equations of non-conventional form
\cite{Nepo01}, while in \cite{Nepo02} he derived conventional
equations for some special cases of the boundary parameters. Cao et
al. \cite{CaoLSW02} succeeded in formulating a more general Bethe
Ansatz solution of conventional form using the algebraic Bethe Ansatz,
provided that the parameters appearing in the Hamiltonian 
satisfy a certain constraint. The same equations were subsequently also
found by Nepomechie \cite{Nepo03}. 

This solution was studied numerically and analytically in 
\cite{NepoR03,AhnN03}, and it was noted that it did not describe the
full spectrum for all parameter values satisfying the constraint. In
an addendum to \cite{NepoR03} a simple extension of the original Bethe
Ansatz solution was conjectured on numerical grounds that does
describe the complete spectrum, again provided the constraint is
satisfied. In this paper we will give a derivation of the complete
spectrum of the XXZ chain, also with the condition that a particular
constraint has to be satisfied. For even system sizes this constraint
is a special case of the constraint considered by the authors above,
but it is different for odd $L$. 

We use the fact that the XXZ spin chain can be described in terms
of the Temperley-Lieb algebra with boundary generators. In the
diagrammatic representation of this algebra the Hamiltonian of the
spin chain is that of a loop model. We show that there exist a simple
eigenstate in this loop model that serves as a pseudo vacuum to set up
a coordinate Bethe Ansatz calculation, as originally used by Bethe
\cite{Bethe31}. We thus are able to derive the Bethe Ansatz equations,
some of which were conjectured in \cite{NepoR03}. The approach
followed in this paper is a generalisation of the method described in
\cite{Levy90,Levy91a,MartS94a}.

For most sectors of the loop Hamiltonian, whose spectrum is larger
than that of the spin chain, there is no additional constraint on the
parameters. However, for the sector containing the spin chain
spectrum, we do need to restrict ourselves with a constraint similar
to the one found in \cite{CaoLSW02} and \cite{Nepo03}. This constraint
is very natural in terms of the loop model since it preserves a
${\mathbb Z}_2$ symmetry (see Section \ref{se:BAintro}). When the
constraint is not satisfied, the symmetry is broken 
and one loses a conservation law which introduces additional
difficulties in formulating a Bethe Ansatz. We hope however that our
formulation offers new insights for finding the general solution of
the spectrum.

In Section \ref{se:algebra} we introduce the Temperley-Lieb loop model
with boundary generators, and show its relation to the quantum
spin-$1/2$ XXZ chain with non-diagonal boundary conditions. In Section
\ref{se:BAintro} we show that the loop model has a simple eigenstate
which we can use as a pseudo vacuum to setup a Bethe Ansatz
calculation. In this section we also identify the pseudo particles
that label the sectors of the Hamiltonian. Before presenting a
detailed derivation in Sections \ref{se:closed}, \ref{se:mixed} and
\ref{se:open}, we state our main result in Section \ref{se:result}.

\sectiona{Algebra and representations}
\label{se:algebra}
\subsection{The Temperley Lieb algebra with boundaries}
\label{se:TL}
The Open Temperley-Lieb (OTL) algebra $T_L^{\rm O}$ with generators $e_x$
($x=1,\ldots,L-1$) and boundary generators $f_+$ and $f_-$ is defined
as follows \cite{MitraNGB03},
\be
\begin{array}{ll}
\dps e_x^2 = t e_x & \dps f_-^2 = s_- f_- \\
\dps e_x e_{x\pm 1} e_x = e_x & \dps f_+^2 = s_+ f_+ \\[2mm]
\dps e_xe_y = e_y e_x\quad {\rm for}\; |x-y| \leq 2 & f_-f_+=f_+f_-\\
\dps e_1f_-e_1= e_1 & e_xf_- =f_-e_x \quad {\rm for}\; x>1\\
\dps e_{L-1}f_+e_{L-1}= e_{L-1} & e_xf_+ =f_+e_x \quad {\rm for}\; x<L-1
\end{array} 
\ee
This algebra is infinite dimensional. It can be made finite
dimensional by introducing
\be
\renewcommand{\arraystretch}{3}
\begin{array}{ll}
\dps I_{2n} = \prod_{x=0}^{n-1} e_{2x+1} & \dps J_{2n} = f_- \prod_{x=1}^{n-1}
e_{2x}\; f_+\\
\dps I_{2n+1} = f_-\prod_{x=1}^{n} e_{2x} & \dps J_{2n+1} =
\prod_{x=0}^{n-1} e_{2x+1}\; f_+.
\end{array}
\label{eq:idempot}
\ee
and imposing the following relations,
\be
I_LJ_LI_L=b I_L,\quad J_L I_L J_L = b J_L,
\label{eq:TLb}
\ee
both for $L$ even and $L$ odd. For later convenience, we introduce
here two sub-algebras. The first one is the original Temperley-Lieb
algebra, introduced in \cite{TempL71}, and formed by the generators 
$e_x$ ($x=1,\ldots,L-1$). As it does not contain boundary generators we
will call it the Closed TL algebra and denote it here by 
$T_L^{\rm C}$. The second sub-algebra we consider is the Mixed TL
algebra containing only the boundary generator $f_-$. We denote it by
$T_L^{\rm M}$, and it is formed by the generators $e_x$
($x=1,\ldots,L-1$) and $f_-$. The algebra $T_L^{\rm M}$ is the blob
algebra introduced in \cite{Levy91a,Levy91b,MartS94b}. We further note
that the relation (\ref{eq:TLb}) does not apply to the algebras $T_L^{\rm C}$
and $T_L^{\rm M}$, but only to the full algebra $T_L^{\rm O}$.   

In this paper we will consider the operators 
\bea
H^{\rm C}_L &=& \sum_{x=1}^{L-1} e_x, \label{eq:hamC}\\
H^{\rm M}_L &=& a_-f_- + \sum_{x=1}^{L-1} e_x,\label{eq:hamM}\\
H^{\rm O}_L &=& a_-f_- + a_+f_+ + \sum_{x=1}^{L-1} e_x.
\label{eq:hamO}
\eea
These operators correspond in certain representations to the quantum
spin-$1/2$ XXZ Hamiltonian with various types of open boundary
conditions, see Section \ref{se:XXZrep}. The main result of this paper is the
diagonalisation of these operators in a different representation,
namely the loop representation which we will explain now.

\subsection{The loop representation}
\label{se:looprep}
Instead of the algebra it is convenient to use the graphical
representation of $T_L^{\rm O}$,
\be
e_x \quad =\quad
\begin{picture}(130,20)
\put(0,0){\epsfxsize=130pt\epsfbox{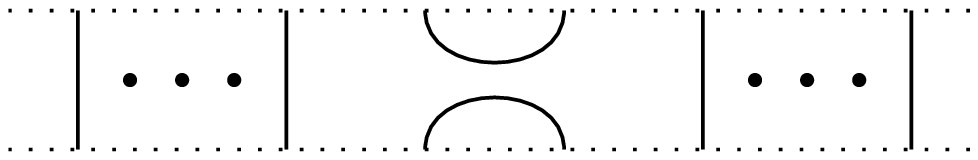}}
\put(51,-10){$\sstyle x$}
\put(67,-10){$\sstyle x+1$}
\end{picture}\;,
\label{eq:monoid}
\ee
and
\vskip6pt
\be
\begin{picture}(240,20)
\put(0,10){$f_-=$}
\put(30,0){\epsfxsize=70pt\epsfbox{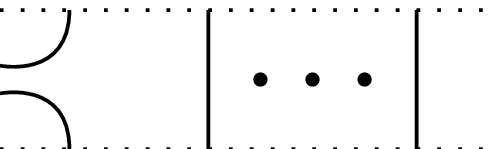}}
\put(140,10){$f_+=$}
\put(170,0){\epsfxsize=70pt\epsfbox{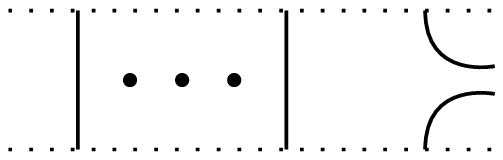}}
\end{picture}
\label{eq:f1-fn}
\ee
Multiplication of two words in the algebra corresponds to putting one
word below the other and merging the loops lines. For example, the
relations $e_x^2=t e_x$ and $e_xe_{x+1}e_x=e_x$ graphically read
\be
\begin{picture}(120,40)
\put(0,0){\epsfxsize=40pt\epsfbox{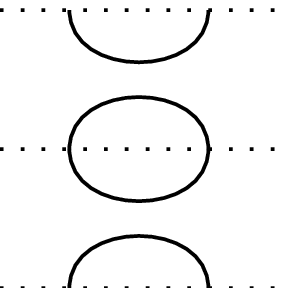}}
\put(50,18){$=\;\;t$}
\put(80,10){\epsfxsize=40pt\epsfbox{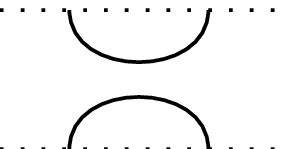}}
\end{picture}
\label{eq:e^2}
\ee
\vskip2mm
\be
\begin{picture}(150,60)
\put(0,0){\epsfxsize=60pt\epsfbox{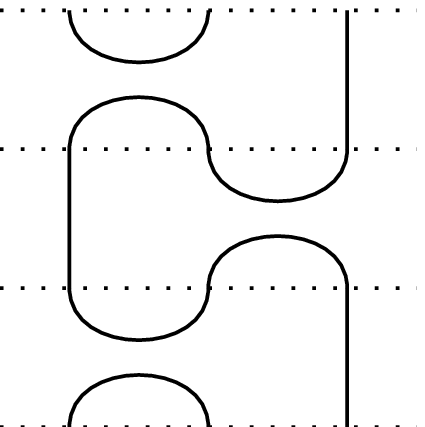}}
\put(70,28){$=$}
\put(90,20){\epsfxsize=60pt\epsfbox{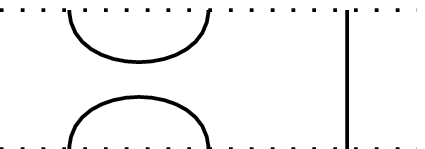}}
\end{picture}
\ee
The loop representation is obtained by filtering the algebra by
quotients of the left ideals generated by the (unnormalized)
idempotents $I_n$ and $J_n$. This procedure is described in detail in
\cite{Martin91}, here we will explain this representation pictorially.
A crucial property of the loop representation is that it has a heighest
weight state, which is represented graphically by $L$ vertical lines,
i.e. it is the state in which no sites are connected by loop lines. We
will take the view that in this case the sites are ``connected to
infinity''. Let us denote this highest weight state by $\ket{}$. All
other states in the loop representation can be obtained by applying
the words of the algebra to this highest weight state. The pictures
corresponding to these other states are obtained by placing the
picture of the word underneath that of the highest weight state and
removing disconnected parts from the top of the combined picture. For
example, the state corresponding to $e_x \ket{}$ will be represented
by the bottom half of the picture in (\ref{eq:monoid}),    
\be
e_x\ket{}\quad = \quad \ket{x} \quad = \quad
\begin{picture}(140,10)
\put(0,0){\epsfxsize=140pt\epsfbox{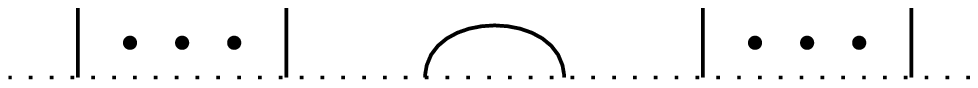}}
\put(56,-10){$\sstyle x$}
\put(72,-10){$\sstyle x+1$}
\end{picture}\;,
\label{eq:eiT}
\ee
\vskip6pt
\noindent
where we have denoted by $\ket{x}$ the state in which sites $x$
and $x+1$ are connected to each other while all other sites are
connected to infinity. 

It is helpful to understand the action of $e_{x+1}$ on
$\ket x $, which is given by the following pictures,
\bea
e_{x+1}e_x\ket{} \quad &=&\quad 
\begin{picture}(160,30)
\put(0,0){\epsfxsize=160pt\epsfbox{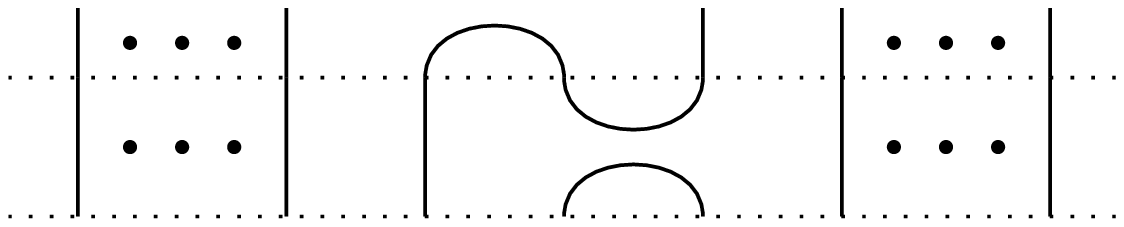}}
\put(56,-10){$\sstyle x$}
\put(72,-10){$\sstyle x+1$}
\put(92,-10){$\sstyle x+2$}
\end{picture}\nonumber\\[14pt]
&=& \quad
\begin{picture}(160,10)
\put(0,0){\epsfxsize=160pt\epsfbox{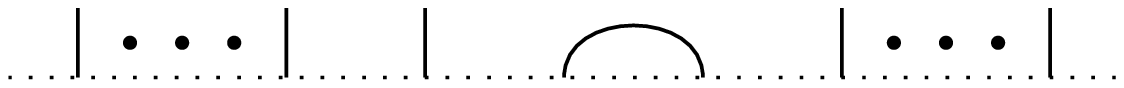}}
\put(56,-10){$\sstyle x$}
\put(72,-10){$\sstyle x+1$}
\put(92,-10){$\sstyle x+2$}
\end{picture} \quad = \quad \ket{x+1}.
\label{eq:eip1T}
\eea
\vskip6pt
\noindent
A slightly more complicated example is given by the picture
representing $e_{x+1}e_{x+2}e_x\ket{}$, 
\be
e_{x+1}e_{x+2}e_x\ket{}\quad = \quad
\begin{picture}(170,15)
\put(0,0){\epsfxsize=170pt\epsfbox{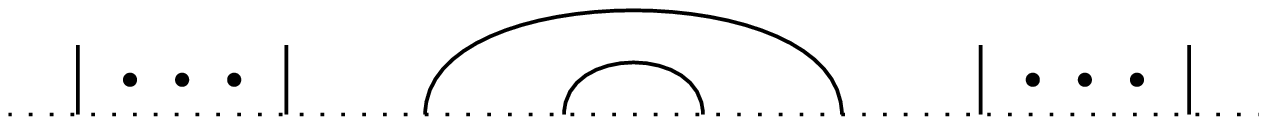}}
\put(52,-10){$\sstyle x$}
\put(106,-10){$\sstyle x+3$}
\end{picture}\;.
\label{eq:eiTcomp}
\ee
\vskip6pt 
\noindent
Analogous pictures are obtained when the boundary generators $f_\pm$
are involved. In particular, the first relation in (\ref{eq:TLb}) for
$L=2$ reads graphically,
\be
\begin{picture}(246,50)
\put(0,20){$e_1f_-f_+e_1\ket{}\;=$}
\put(80,0){\epsfxsize=40pt\epsfbox{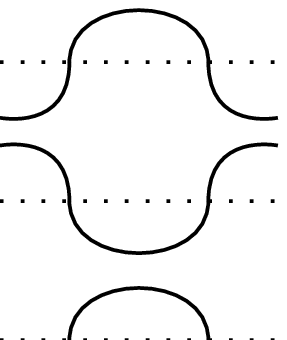}}
\put(130,20){$=\; b\;$}
\put(156,20){\epsfxsize=40pt\epsfbox{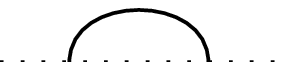}}
\put(206,20){$=\; b\; e_1\ket{}$.}
\end{picture}
\label{eq:ef1f2e}
\ee
The parameter $b$ in (\ref{eq:TLb}) thus is the weight of the two
removed horizontal loop lines. Note that the state $e_1 f_-f_+\ket{}$
has the same lower part of the picture as $e_1\ket{}$, but contains
only one horizontal loop line. This cannot be removed using the
algebraic relations, but in analogy with the above we do have 
for $L=2$ that 
\be
f_-f_+e_1 f_-f_+\ket{}= b f_-f_+ \ket{} .
\ee
In general we find two isomorphic invariant subspaces for $b\neq 0$,
namely the ones generated by $I_L \ket{}$ and $J_L \ket{}$. Because
the spectrum of $H_L^{\rm O}$ is identical on them, even when $b=0$,
we will discard the latter subspace. Hence, when considering 
cases where all loop lines are connected we will always work
exclusively in the one generated by $I_L \ket{}$. In this way we construct the
loop representations for the algebras $T_L^{\rm O}$, $T_L^{\rm M}$ and
$T_L^{\rm C}$ and we denote the representation spaces by $V_L^{\rm
C}$, $V_L^{\rm M}$ and $V_L^{\rm O}$ respectively. It is furthermore
important to note that the loop lines are not allowed to cross. 

Instead of this graphical loop representation we will use a more
convenient typographical notation in the following. If a site $x$ is
connected to a site $y<x$ or to the left boundary, we write a closing
parenthesis ``$)$'' at $x$. If a site $x$ is connected to a site $y>x$
or to the right boundary we write an opening parenthesis ``$($'' at
$x$. If a site is not connected (or connected to infinity) we write a
vertical bar ``$|$'' at $x$. For the sub-algebra $T_2^{\rm M}$ we thus
have the state space,  
\be 
V_2^{\rm M} = {\rm Span} \{ ||, )|, (), ))\}.
\ee
There is a similar identification for $T_2^{\rm O}$ where now loops
can also connect to the right boundary. For example, $|($ denotes a
state where the loop segment on the second site is connected to the
right boundary. The seven basis states of $V_2^{\rm O}$ are
\be
||,\; )|,\; |(,\; (),\; )(,\; )),\; ((.
\ee
We will call a sequence of the three symbols ``$)$'', ``$($'' and
``$|$'' a connectivity.
 
The dimensions of the representation spaces $V_L^{\rm C}$, $V_L^{\rm
M}$ and $V_L^{\rm O}$ can be calculated by counting the number of
sequences of ``$)$'', ``$($'' and ``$|$'' subject to the non-crossing
constraint (for example, ``$|)$'' cannot occur). In the appendix we
find the generating functions for these dimensions. Asymptotically
they are given by
\be
\dim V_L^{\rm C} \approx L^{-1/2}2^L, \quad \dim V_L^{\rm M} = 2^L,
\quad \dim V_L^{\rm O} \approx L^{1/2}2^L.
\ee

\subsection{XXZ representation}
\label{se:XXZrep}

The following representation of the TL algebra makes Hamiltonian
(\ref{eq:hamO}) up to a constant equal to that of the XXZ spin chain with non-diagonal
boundary terms \cite{Nepo03},
\be
e_x = -\frac12\left( \sigma^1_x\sigma^1_{x+1} +
\sigma^2_x\sigma^2_{x+1} + \ch\, \eta \;\sigma^3_x\sigma^3_{x+1}
-\sh\,\eta\; (\sigma^3_x - \sigma^3_{x+1})-\ch\,\eta \right) 
\ee
\bea
f_- &=& \frac{1}{2\rho_-} \left(\sh(\alpha_-+\beta_-) \sigma^3_1
+ \ch\,\theta_-\; \sigma^1_1 + \i\, \sh\,\theta_-\; \sigma^2_1 +
\ch(\alpha_-+\beta_-) \right), \\
f_+ &=& \frac{1}{2\rho_+} \left(-\sh(\alpha_++\beta_+) \sigma^3_L
+ \ch\,\theta_+\; \sigma^1_L + \i\, \sh\,\theta_+\; \sigma^2_L +
\ch(\alpha_++\beta_+) \right), 
\eea
where
\be
\rho_\mp=\ch(\alpha_\mp+\beta_\mp+\eta),
\ee
and $\sigma^i_x$ represents the $i$th Pauli matrix at site $x$. The
parameters $t$, $a_\mp$ and $s_\mp$ are given by 
\bea
t &=& 2\ch\,\eta, \nonumber\\
a_\mp &=& -\sh\,\eta\;\frac{\ch(\alpha_\mp+\beta_\mp+\eta)}
{\sh\,\alpha_\mp\ch\,\beta_\mp},\\
s_\mp &=&
\frac{\ch(\alpha_\mp+\beta_\mp)}{\ch(\alpha_\mp+\beta_\mp+\eta)},
\nonumber
\eea
while the expression for $b$ depends on the parity of the system size $L$,
\be
b =
\renewcommand{\arraystretch}{2.2}
\left\{ \begin{array}{ll}
\dps \frac{\ch(\eta+\alpha_-+\alpha_++\beta_-+\beta_+)-\ch(\theta_--\theta_+)}
     {2\ch(\alpha_-+\beta_-+\eta)\,\ch(\alpha_++\beta_++\eta)}\quad &
     {\rm for\; even}\;L,\\
\dps\frac{\ch(\alpha_--\alpha_++\beta_--\beta_+)+\ch(\theta_--\theta_+)}
     {2\ch(\alpha_-+\beta_-+\eta)\,\ch(\alpha_++\beta_++\eta)} &
     {\rm for\; odd}\;L.
\end{array}\right.
\ee

The dimension of this spin-$1/2$ representation is $2^L$. It is
important to note that the dimension of $V_L^{\rm O}$ is larger than
this. However, $\dim V_L^{\rm M}=2^L$ and this representation
seems to be equivalent to XXZ.  

The Hamiltonian $H_L^{\rm O}$, given in (\ref{eq:hamO}), is related to
the Hamiltonian $\mathcal{H}$ given in \cite{NepoR03} by
\be
\renewcommand{\arraystretch}{2.0}
\begin{array}{rcl}
\dps H_L^{\rm O} &=& \dps {}-\mathcal{H} +\frac14(L-1)t +\frac12(a_-s_-+a_+s_+) \\
&=& \dps{}-\mathcal{H} +\frac12(L-1)\ch\, \eta - \frac12 \sh\,\eta\;
(\cth\, \alpha_- + \th\, \beta_- +\cth\, \alpha_+ + \th\, \beta_+).
\end{array}
\ee

\sectiona{Pseudo particles}
\label{se:BAintro}

In the representations on the spaces $V_L^{\rm C}$,  $V_L^{\rm M}$ and
$V_L^{\rm O}$ the Hamiltonians (\ref{eq:hamC})-(\ref{eq:hamO}) have a
block triangular structure: connections can be created but cannot be
undone. This implies that for each of the Hamiltonians
(\ref{eq:hamC})--(\ref{eq:hamO}) the state $\bra{}$, dual to $\ket{}$,
is a {\em left} eigenvector with eigenvalue $0$ (since there are no
nonzero matrix elements giving transitions {\em to} this state). More generally,
although the Hamiltonians are not block diagonal, the number of connected sites 
(either to another site or to the boundary) is still a good quantum
number to characterize the spectrum and its corresponding set of {\em
left} eigenvectors, which are elements of the dual spaces $V_L^{\rm C*}$,
$V_L^{\rm M*}$ and $V_L^{\rm O*}$. We will denote the basis states in the
dual space $V_L^*$ by the same connectivities as their dual states in $V_L$ (recall
that a connectivity is a sequence of the three symbols ``$)$'', ``$($'' and 
``$|$'') . Let $\bra c$ denote the basis state corresponding to the
connectivity $c$. A vector $\bra\psi \in V_L^*$ then is decomposed on the dual basis as
\begin{equation}
\bra\psi = \sum_c \psi(c) \bra c,
\end{equation}
and the left eigenvalue equation we want to solve reads
\begin{equation}
\Lambda \bra\psi = \bra\psi H,
\quad \Leftrightarrow\quad  \Lambda \psi(c) = \sum_{c'} \psi(c') H_{c'c}.
\label{eq:eigval_left}
\end{equation}

In the following sections we will calculate the spectra by formulating a 
coordinate Bethe Ansatz for the left eigenvectors of each of the
Hamiltonians. Here we want to briefly discuss the main ingredients.
The trivial eigenstate will be used as a pseudo vacuum
for the Bethe Ansatz. In setting up the Bethe Ansatz calculation we
need to identify ``pseudo particles'' or ``elementary excitations'' on
this pseudo vacuum. We will introduce the notion of pseudo particle for
the space $V_L$. It is then easy to translate this notion to the space
$V^*_L$ by duality.

A link between $x$ and $x+1$, i.e the graphical
representation of $\ket x $ in (\ref{eq:eiT}), will be such a pseudo
particle. Since in this representation $e_{x\pm1}\ket x =
\ket{x\pm1}$, the bulk Hamiltonian (\ref{eq:hamC}) describes
hopping of the pseudo particle. It furthermore contains a diagonal
term ($e_{x}\ket x =t\ket x $) and creates pseudo particles at
other sites through the action of $e_y$ with $y\neq x-1,x$ or
$x+1$. For example, a state with two pseudo particles at $x$ and $y$
respectively is given by
\be
e_x\ket y \quad = \quad \ket{x,y}\quad = \quad
\begin{picture}(210,10)
\put(0,0){\epsfxsize=210pt\epsfbox{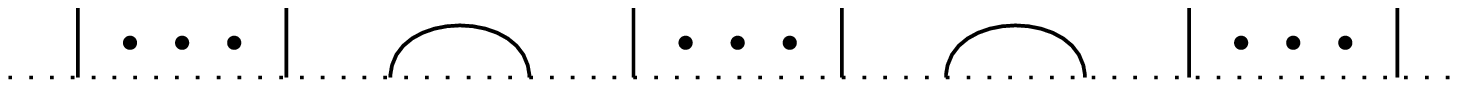}}
\put(51,-10){$\sstyle x$}
\put(67,-10){$\sstyle x+1$}
\put(132,-10){$\sstyle y$}
\put(148,-10){$\sstyle y+1$}
\end{picture}
\label{eq:eiejT}
\ee
\vskip6pt
\noindent
The states with well separated pseudo particles form just a small
subset of all possible states. All the other states however will be
considered as bound states of pseudo particles. For example, we
consider a state of the form (\ref{eq:eiTcomp}) as a bound state of
two pseudo particles.  

In $V_L^{\rm C}$ we only have states where sites are connected
pairwise. When boundaries are present, sites do not have to be paired to be
connected, they can be connected to the left or right boundary. The
number of connected sites can therefore take on odd values. In the
following we will take the number of connected sites as our quantum
number labeling the different sectors of the Hamiltonians.

Anticipating our results, we note here that when $b=0$ in
(\ref{eq:TLb}), the Hamiltonian $H_L^{\rm O}$ possesses an additional
${\mathbb Z}_2$ symmetry. From (\ref{eq:ef1f2e}), and more generally
from
\be
I_L J_L I_L \ket{}= b\, I_L\ket{},
\ee
it follows that $H_L^{\rm O}$ is block triangular when $b=0$ with
respect to the subspaces with an odd or even number of sites connected
to the left boundary. It is clear that this symmetry is always present
in sectors where the number of connected sites is less than the system
size.  

\sectiona{Results}
\label{se:result}
Before giving a detailed derivation of the Bethe Ansatz for the
Temperley-Lieb loop model, we would like to state our main result
first. For convenience, we define the following two functions,
\be
\phi_\pm(u,\epsilon) =
\frac{\sh(\eta/2+\epsilon\alpha_\pm-u)\,\ch(\eta/2+\epsilon\beta_\pm-u)}
    {\sh(\eta/2+\epsilon\alpha_\pm+u)\,\ch(\eta/2+\epsilon\beta_\pm+u)},
\ee
where $\epsilon^2=1$.
The eigenvalues of the loop model $H_L^{\rm O}$ for sectors labeled by
$n$ are given by,
\begin{itemize}
\item $n$ even, $n<L$ 

For these sectors, the spectrum is described by the expressions
\be
\renewcommand{\arraystretch}{1.4}
\begin{array}{rcl}
\dps \Lambda_0 &=& \dps \sum_{j=1}^{n/2} \frac{\sh^2\eta}{\sh(\eta/2-u_j)
  \,\sh(\eta/2+u_j)}, \\
\dps \Lambda_1 &=& \dps \sum_{j=1}^{n/2-1} \frac{\sh^2\eta}{\sh(\eta/2-v_j)
  \,\sh(\eta/2+v_j)} \\
&& \dps {} - \sh\,\eta\; (\cth\, \alpha_- + \th\, \beta_- +
  \cth\, \alpha_+ + \th\, \beta_+),
\end{array}
\label{eq:BAneven1}
\ee
where the complex numbers $u_i$ and $v_i$ are solutions of the equations
\be
\left( \frac{\sh(\eta/2 -u_i)} {\sh(\eta/2 +u_i)} \right)^{2L} =
\phi_-(u_i,+) \phi_+(u_i,+)
\prod_{\stackrel{\scriptstyle j=1}{j \neq i}}^{n/2}
\frac{\sh(u_i-u_j+\eta) \,\sh(u_i+u_j+\eta)} {\sh(u_i-u_j-\eta)
  \,\sh(u_i+u_j-\eta)},
\label{eq:BAneven2}
\ee
and
\be
\left( \frac{\sh(\eta/2 -v_i)} {\sh(\eta/2 +v_i)} \right)^{2L} =
\phi_-(v_i,-) \phi_+(v_i,-)
\prod_{\stackrel{\scriptstyle j=1}{j \neq i}}^{n/2-1}
\frac{\sh(v_i-v_j+\eta) \,\sh(v_i+v_j+\eta)} {\sh(v_i-v_j-\eta)
  \,\sh(v_i+v_j-\eta)}.
\label{eq:BAneven3}
\ee
These are the Bethe Ansatz equations conjectured in the addendum of
\cite{NepoR03} for the special case of $k=1$. A similar division into
two different functional forms was found recently in the ground state
energy for the two-boundary sine-Gordon model \cite{CauxSS02}.

\item $n$ odd, $n<L$ 

For these sectors, the spectrum is described by the expressions
\be
\renewcommand{\arraystretch}{1.4}
\begin{array}{rcl}
\dps \Lambda_0 &=& \dps \sum_{j=1}^{(n-1)/2} \frac{\sh^2\eta}{\sh(\eta/2-u_j)
  \,\sh(\eta/2+u_j)} - \sh\,\eta\; (\cth\, \alpha_- + \th\, \beta_- ), \\
\dps \Lambda_1 &=& \dps \sum_{j=1}^{(n-1)/2} \frac{\sh^2\eta}{\sh(\eta/2-v_j)
  \,\sh(\eta/2+v_j)} - \sh\,\eta\; (\cth\, \alpha_+ + \th\, \beta_+),
\end{array}
\label{eq:BAnodd1}
\ee
where the complex numbers $u_i$ and $v_i$ are solutions of the equations
\be
\left( \frac{\sh(\eta/2 -u_i)} {\sh(\eta/2 +u_i)} \right)^{2L} =
\phi_-(u_i,-) \phi_+(u_i,+)
\prod_{\stackrel{\scriptstyle j=1}{j \neq i}}^{(n-1)/2}
\frac{\sh(u_i-u_j+\eta) \,\sh(u_i+u_j+\eta)} {\sh(u_i-u_j-\eta)
  \,\sh(u_i+u_j-\eta)},
\label{eq:BAnodd2}
\ee
and
\be
\left( \frac{\sh(\eta/2 -v_i)} {\sh(\eta/2 +v_i)} \right)^{2L} =
\phi_-(v_i,+) \phi_+(v_i,-)
\prod_{\stackrel{\scriptstyle j=1}{j \neq i}}^{(n-1)/2}
\frac{\sh(v_i-v_j+\eta) \,\sh(v_i+v_j+\eta)} {\sh(v_i-v_j-\eta)
  \,\sh(v_i+v_j-\eta)}.
\label{eq:BAnodd3}
\ee

\item $n=L$ even

In this sector, the spectrum of $H_L^{\rm O}$ is still given
by (\ref{eq:BAneven1})-(\ref{eq:BAneven3}) provided that
\be
b=0 \quad \Leftrightarrow \quad \ch (\eta +\alpha_-+\beta_- +
\alpha_++\beta_+) -\ch (\theta_--\theta_+)=0,
\label{eq:evenconstr}
\ee
where $b$ is the parameter appearing in (\ref{eq:TLb}).
The spectrum of $H_L^{\rm O}$ in this sector with the additional
constraint is identical to that of the XXZ spin chain with an even
number of spins, as conjectured in the addendum of \cite{NepoR03}. 

\item $n=L$ odd

In this case, the spectrum of $H_L^{\rm O}$ is still given
by (\ref{eq:BAnodd1})-(\ref{eq:BAnodd3}) provided that
\be
b=0 \quad \Leftrightarrow \quad \ch (\eta +\alpha_-+\beta_- -
\alpha_+-\beta_+) +\ch (\theta_--\theta_+)=0.
\label{eq:oddconstr}
\ee
The spectrum of $H_L^{\rm O}$ in this sector is identical to that of
the XXZ spin chain with an odd number of spins. Note that while the
constraint (\ref{eq:evenconstr}) for even system sizes is the same as
that obtained in \cite{Nepo03} for $k=1$, the constraint
(\ref{eq:oddconstr}) is similar but not identical to the one in
\cite{Nepo03} with $k=0$. That constraint corresponds to $b=s_-s_+$, a
case that is not discussed in this paper (see however Section
\ref{se:n=L=1}). 

\end{itemize}
In the following sections we give a detailed derivation of these
results. As an intermediate result, we also obtain the spectrum of
$H_L^{\rm M}$, which is closely related to a Hamiltonian diagonalised
by Bethe Ansatz in \cite{DoikM02}.

\sectiona{Closed boundaries}
\label{se:closed}

In this section we will diagonalise the Hamiltonian (\ref{eq:hamC}) on
the space $V_L^{\rm C*}$. In the XXZ representation this Hamiltonian
corresponds to the quantum group symmetric Hamiltonian, which has
diagonal, spin conserving, boundary fields. The Bethe Ansatz solution
of the spectrum of this Hamiltonian is well known
\cite{Sklya88,AlcaBBBQ87}. We rederive it here in terms of the
loop model which will be a useful exercise for when we introduce
boundary fields in later sections.

As explained above, a sector of $H_L^{\rm C}$ is labeled by an integer
$n$, denoting the number of connected sites. For closed boundaries $n$
will be even since sites cannot be connected to the boundary, only to
each other. 
\subsection{$n=0$}

Consider the configuration with all sites disconnected, i.e. $\bra{} =
|\cdots|$. We will try to solve (\ref{eq:eigval_left}) with the trial
function $\bra{\psi_0}$ given by,
\be
\bra{\psi_0} = \bra{} .
\label{eq:Cn=0trial}
\ee
Plugging $\bra{\psi_0}$ into (\ref{eq:eigval_left}) we obtain
\be
\Lambda \bra{} =0,
\label{eq:Cn=0a1}
\ee
and immediately obtain that $\bra{\psi_0}$ is a left eigenvector with eigenvalue
\begin{equation}
\Lambda=0,
\end{equation}
as asserted in Section \ref{se:BAintro}. Hence we can use $\bra{}$
as a pseudo vacuum.

\subsection{$n=2$}

Here and in the following we will consider a trial eigenvector
$\bra{\psi_n}$, which is a linear combination of states whose
coefficients for states with more than $n$ connected sites are
zero. The coefficients of $\bra{\psi_n}$ are denoted by $\psi_n(c)$,
where $c$ labels the connectivity. For $n=2$, we thus take a linear
combination of the pseudo vacuum, with coefficient $\psi_2()$, and states with a
pseudo particle at site $x$, i.e. states of the form
$\bra{x}=|\cdots|()|\cdots |$, with nonzero coefficients $\psi_2(x)$. Now
we will try to solve (\ref{eq:eigval_left}) with the trial function
$\bra{\psi_2}$ given by, 
\be
\bra{\psi_2} = \psi_2()\bra{} + \sum_{x=1}^L \psi_2(x) \bra{x}.
\ee
Plugging $\bra{\psi_2}$ into (\ref{eq:eigval_left}) and equating each
component to zero the eigenvalue equation is equivalent to,
\be
\renewcommand{\arraystretch}{1.4}
\begin{array}{rcl}
\dps \Lambda \psi_2() &=& \dps \sum_{x=1}^{L-1} \psi_2(x),\\
\dps \Lambda \psi_2(x) &=& \dps t \psi_2(x) + \psi_2(x+1) + \psi_2(x-1)\qquad
{\rm for}\;\; 1<x<L-1,
\end{array}
\label{eq:Cn=2a}
\ee
together with the following equations at the boundaries $x=1$ and $x=L-1$,
\be
\renewcommand{\arraystretch}{1.2}
\begin{array}{rcl}
\dps \Lambda \psi_2(1) &=& \dps t \psi_2(1) + \psi_2(2),
\\
\dps \Lambda \psi_2(L-1) &=& \dps t \psi_2(L-1) + \psi_2(L-2).
\end{array}
\label{eq:Cn=2c}
\ee
We first note that due to the block triangular structure
of the Hamiltonian, the coefficient $\psi_2()$ only appears in
the first line of (\ref{eq:Cn=2a}), which we therefore regard as defining
$\psi_2()$.\footnote{We need here that $\Lambda \neq 0$. Here and in the
following we will not digress into such isolated singularities.} The
remaining equations can be satisfied by the Ansatz,   
\be
\psi_2(x) = f(x),
\label{eq:Cn=2Ans}
\ee
where $f$ is the function
\be
f(x) = A^+ z^x + A^- z^{-x}.
\ee
The parameters $A^\pm$ and $z$ are complex numbers to be determined
later. We now find from (\ref{eq:Cn=2a}) that
\be
\Lambda = t + z+z^{-1},
\ee
and in addition the consistency relations arising from the boundary
equations (\ref{eq:Cn=2c}),
\be
f(0) = 0,\qquad f(L) = 0.
\label{eq:Cn=2e}
\ee
These consistency relations determine the ratio of the amplitudes $A^\pm$ and the
complex number $z$. In this sector therefore the spectrum is given by
the solutions of
\be
\Lambda = \lambda(z),\qquad z^{2L} = 1,
\ee
where we have defined $\lambda$ for future convenience,
\be
\lambda(z) = t + z+z^{-1}.
\label{eq:lamdef}
\ee

\subsection{$n=4$}
\label{se:Cn=4}
Let us denote by $\psi_4(x_1,x_2)$ the coefficient for states of the
form $|\cdots|()|\cdots|()|\cdots|$, with ``well separated'' pseudo
particles at positions $x_1$ and $x_2$, i.e. $x_2>x_1+1$. We furthermore denote the
coefficient for states of the form $|\cdots|(())|\cdots|$ with
``merged'' pseudo particles by $\varphi_4(x)$. As a trial vector to
solve the eigenvalue equation (\ref{eq:eigval_left}) we consider a
vector with nonzero components $\psi_4(c)$ in all sectors $m$ of $V_L^{\rm
C*}$ with less than four connected sites. Aside from the states with
$m=4$ mentioned above, this trial vector contains also the states
$\ket x$ and $\ket{}$ (with coefficients $\psi_4(x)$ and $\psi_4()$)
in the $m=2$ and $m=0$ sectors respectively. In
analogy with the case $n=2$ we can write down the bulk eigenvalue
equation component wise. However, we will only look at the equations
relating the eigenvector elements $\psi_4(c)$ in the $m=4$ sector to each
other. The equations relating elements $\psi_4(c)$ for $m<4$ to those
of $m=4$ are similar to (\ref{eq:Cn=2a}) and can be regarded as
definitions of $\psi_4(c)$ for $m<4$. We will therefore not write them
down explicitly. The remaining equations are, 
\be
\renewcommand{\arraystretch}{1.2}
\begin{array}{rcl}
\dps \Lambda \psi_4(x_1,x_2) &=& \dps \psi_4(x_1-1,x_2) +\psi_4(x_1+1,x_2) +
\psi_4(x_1,x_2-1) \\
&&{}\dps + \psi_4(x_1,x_2+1) + 2t\psi_4(x_1,x_2)\qquad
{\rm for}\;\; x_2>x_1+2,\\
\dps \Lambda \psi_4(x,x+2) &=& \dps \psi_4(x-1,x+2) +\varphi_4(x) + \psi_4(x,x+3) +
2t\psi_4(x,x+2), \\
\dps \Lambda \varphi_4(x) &=& \dps 2\psi_4(x,x+2) +\psi_4(x-1,x+1) +
\psi_4(x+1,x+3) \\
&&{}\dps + t\varphi_4(x)\qquad {\rm for}\;\; 1<x<L-3. 
\end{array}
\label{eq:Cn=4e}
\ee
As before, these are modified at the boundary to,
\be
\renewcommand{\arraystretch}{1.2}
\begin{array}{rcl}
\dps \Lambda \psi_4(1,x)  &=& \dps \psi_4(2,x) + \psi_4(1,x-1) + \psi_4(1,x+1) +
2t\psi_4(1,x), \\
\dps \Lambda \psi_4(x,L-1)  &=& \dps \psi_4(x-1,L-1) + \psi_4(x+1,L-1)
+ \psi_4(x,L-2)\\ 
&&{}\dps + 2t\psi_4(x,L-1),\\
\dps \Lambda \psi_4(1,3) &=& \dps \psi_4(1,4) + \varphi_4(1) +
2t\psi_4(1,3), \\
\dps \Lambda \psi_4(L-3,L-1) &=& \dps \psi_4(L-4,L-1) + \varphi_4(L-3) +
2t\psi_4(L-3,L-1),\\
\dps \Lambda \varphi_4(1) &=& \dps 2\psi_4(1,3) + \psi_4(2,4) +
t\varphi_4(1),\\
\dps \Lambda \varphi_4(L-3) &=& \dps 2\psi_4(L-3,L-1) + \psi_4(L-4,L-2) +
t\varphi_4(L-3). 
\end{array}
\label{eq:Cn=4k}
\ee
The equations (\ref{eq:Cn=4e}) and (\ref{eq:Cn=4k}) can be solved with the
following Ansatz,
\be
\psi_4(x_1,x_2) = f(x_1,x_2),
\ee
where
\be
f(x_1,x_2)=\sum_{\pi\in S_2} \sum_\sigma A_{\pi_1\pi_2}^{\sigma_1\sigma_2}
z_{\pi_1}^{\sigma_1 x_1} z_{\pi_2}^{\sigma_2 x_2}.
\label{eq:Cn=2Ansatz}
\ee
Here $S_n$ is the group of permutations of $n$ integers $\{1,2,\dots ,n\}$ and
$\pi: \{1,2,\dots , n\}\mapsto \{\pi_1,\pi_2,\dots ,\pi_n\}$ is a 
particular permutation. The sum over $\sigma$ denotes a sum over all signs
$\sigma_1 = \pm 1$ and $\sigma_2 = \pm 1$. The four amplitudes
$A_{12}^{++}$, $A_{21}^{++}$, $A_{12}^{--}$ and $A_{21}^{--}$ as well
as the two complex numbers $z_1$ and $z_2$ are to be determined
later. The form of $\varphi_4(x)$ will be also derived from the eigenvalue
equations. 

From the bulk relations (\ref{eq:Cn=4e}) we find
\be 
\Lambda = \sum_{j=1}^2 \lambda(z_j),
\ee
where $\lambda$ is defined in (\ref{eq:lamdef}), and we also find
\bea
\varphi_4(x) &=& f(x,x+1)+f(x+1,x+2), 
\label{eq:Cn=4phi}\\
0 &=& f(x,x) + 2f(x+1,x+1) + f(x+2,x+2) \nonumber\\ 
&& {}+ t(f(x,x+1)+f(x+1,x+2)).
\label{eq:Cn=4scat}
\eea
Equation (\ref{eq:Cn=4phi}) defines the coefficient $\varphi_4(x)$ and (\ref{eq:Cn=4scat})
is satisfied if,
\be
\frac{A_{\pi_1\pi_2}^{\sigma_1\sigma_2}}{A_{\pi_2\pi_1}^{\sigma_2\sigma_1}} =
-\frac{S(z_{\pi_2}^{\sigma_2},z_{\pi_1}^{\sigma_1})}
{S(z_{\pi_1}^{\sigma_1},z_{\pi_2}^{\sigma_2})},
\ee
where
\be
S(z,w) = 1 + t w + z w.
\label{eq:Sdef}
\ee
In addition, it can also be shown that the first two of the boundary
relations (\ref{eq:Cn=4k}) are equivalent to,
\be
f(0,x)=0,\qquad f(x,L)=0,
\ee
which can be satisfied by demanding that
\be
z_{\pi_1}^{2\sigma_1 L} = - \frac{A_{\pi_2\pi_1}^{\sigma_2,-\sigma_1}}
{A_{\pi_2\pi_1}^{\sigma_2,\sigma_1}},\quad
1 = - \frac{A_{\pi_1\pi_2}^{-\sigma_1,\sigma_2}}
{A_{\pi_1\pi_2}^{\sigma_1,\sigma_2}}, \label{eq:Cleftbc}
\ee 
for all choices of the signs $\sigma_1=\pm 1$, $\sigma_2=\pm 1$ and both permutations of
$\pi=(1,2)$. It turns out that the remaining equations
do not impose any further constraints and are now automatically
satisfied. Putting everything together we find that
\be 
z_{\pi_1}^{2L} =  - \frac{A_{\pi_2\pi_1}^{+-}}
{A_{\pi_1\pi_2}^{-+}} \frac{A_{\pi_1\pi_2}^{-+}}
{A_{\pi_1\pi_2}^{++}} \frac{A_{\pi_1\pi_2}^{++}}
{A_{\pi_2\pi_1}^{++}} = \frac{S(z_{\pi_1}^{-1},z_{\pi_2})S(z_{\pi_2},z_{\pi_1})}
{S(z_{\pi_2},z_{\pi_1}^{-1})S(z_{\pi_1},z_{\pi_2})},
\label{eq:Cn=4BAE}
\ee
for both permutations of $\pi=(1,2)$. We have found the two-particle
Bethe Ansatz equation of the XXZ spin chain.

\subsection{general $n=2k$}

Because of the integrability of this model, the eigenvalue equations
for sectors containing more than two pseudo particles all reduce to
those for two particles. The  program of the previous sections
can therefore be carried out for larger values of $n$. As a trial function
for a given $n$ we take a vector 
in which all coefficients in sectors higher than $n$ are zero. We also note 
that coefficients of states in sectors lower than $n$ are defined
by the eigenvalue equation in terms of those for $n$. To find the
latter, we assume that the wave function elements
$\psi_{2k}(x_1,\ldots,x_k)$, corresponding to well separated pseudo 
particles, have the following form,
\be
\psi_{2k}(x_1,\ldots,x_k)=\sum_{\pi\in S_k}\sum_\sigma
A_{\pi_1\ldots\pi_k}^{\sigma_1\ldots\sigma_k} \prod_{i=1}^k
z_{\pi_i}^{\sigma_i x_i},
\ee
where, as in (\ref{eq:Cn=2Ansatz}), $S_k$ is the group of permutations
of $k$ integers $\{1,2,\dots ,k\}$ and the sum over $\sigma$ denotes a
sum over all signs $\sigma_i = \pm 1$ ($i=1,\ldots,k$). 
We also assume that the other wave function elements in the same sector can be
consistently calculated from the eigenvalue equation if the amplitudes
satisfy
\be
\frac{A_{\ldots \pi_i \pi_{i+1} \ldots}^{\ldots \sigma_i \sigma_{i+1}
    \ldots}} {A_{\ldots \pi_{i+1} \pi_i \ldots}^{\ldots
    \sigma_{i+1}\sigma_i \ldots}}
=
-\frac{S(z_{\pi_{i+1}}^{\sigma_{i+1}},z_{\pi_i}^{\sigma_i})}
{S(z_{\pi_i}^{\sigma_i},z_{\pi_{i+1}}^{\sigma_{i+1}})}, 
\ee
and
\be
z_{\pi_n}^{2\sigma_n L} = - \frac{A_{\ldots\pi_n}^{\ldots -\sigma_n}}
{A_{\ldots\pi_n}^{\ldots\sigma_n}},\quad
1 = - \frac{A_{\pi_1\ldots}^{-\sigma_1,\ldots}}
{A_{\pi_1\ldots}^{\sigma_1,\ldots}}. \label{eq:Cleftbc_gen}
\ee 
The complex numbers $z_i$ satisfy the obvious generalization of
(\ref{eq:Cn=4BAE}) to arbitrary but even values of $n$,
\be
z_i^{2L} = \prod_{\stackrel{\scriptstyle j=1}{j \neq i}}^{n/2} 
\frac{S(z_i^{-1},z_j)S(z_j,z_i)}{S(z_j,z_i^{-1})S(z_i,z_j)}.
\label{eq:Cn=genBAE}
\ee
Furthermore, the spectrum is given by
\be
\Lambda = \sum_{j=1}^{n/2} \lambda(z_j),
\ee
where we recall that $\lambda$ and $S$ are defined in
(\ref{eq:lamdef}) and (\ref{eq:Sdef}) respectively. Equation
(\ref{eq:Cn=genBAE}) is of course the well known Bethe Ansatz 
equation for the XXZ spin chain with diagonal boundary conditions
\cite{AlcaBBBQ87,Sklya88}. It
is reproduced here from a Bethe Ansatz in the loop representation of
the Temperley-Lieb algebra. Using the well known transformation
\be
z_i = \frac{\sh(\eta/2 +u_i)}{\sh(\eta/2 -u_i)},
\label{eq:z2u}
\ee
the eigenvalue takes on the familiar form,
\be
\Lambda = \sum_{j=1}^{n/2} \frac{\sh^2\eta}{\sh(\eta/2-u_j)
  \,\sh(\eta/2+u_j)}, 
\ee
where the complex numbers $u_i$ are solutions of the equations
\be
\left( \frac{\sh(\eta/2 -u_i)} {\sh(\eta/2 +u_i)} \right)^{2L} =
\prod_{\stackrel{\scriptstyle j=1}{j \neq i}}^{n/2}
\frac{\sh(u_i-u_j+\eta) \,\sh(u_i+u_j+\eta)} {\sh(u_i-u_j-\eta)
  \,\sh(u_i+u_j-\eta)}.
\ee

\sectiona{Mixed boundaries}
\label{se:mixed}

A similar program as for closed boundaries can be carried out for
mixed boundaries, i.e. we can diagonalise $H_L^{\rm M}$ on the space
$V_L^{\rm M*}$. However, now $n$ can take on odd values as well,
representing sectors where an odd number of loop segments are
connected to the left boundary. We will denote by
$\psi_{2k+l}(x_1,\ldots,x_k;y_1,\ldots,y_l)$ a coefficient in the
eigenvector of a state with $k$ well separated pseudo particles at
positions $x_i$ ($i=1,\ldots,k$) and $l$ sites at positions $y_j$
($j=1,\ldots,l$) connected to the left boundary. 
\subsection{$n=0$}

We again start with the state with all sites disconnected, i.e. $\bra\
= |\cdots|$, and find that it is a left eigenstate of $H_L^{\rm M}$ with
eigenvalue $\Lambda=0$.

\subsection{$n=1$}

Next we consider eigenvectors in which the coefficient $\psi_1(;)$ of
the pseudo vacuum is nonzero as well as $\psi_1(;1)$ of the state
$)|\cdots|$ with one site connected to the left boundary. The
eigenvalue equations are 
\be
\renewcommand{\arraystretch}{1.2}
\begin{array}{rcl}
\dps \Lambda \psi_1(;) &=& \dps a_- \psi_1(;1),\\
\dps \Lambda \psi_1(;1) &=& \dps a_- s_- \psi_1(;1).
\end{array}
\label{eq:Mn=1a}
\ee
As before we regard the first line in (\ref{eq:Mn=1a}) as defining $\psi_1(;)$. The
remaining equation is solved by $\Lambda=a_- s_-$.

\subsection{$n=2$}

Now we also consider nonzero coefficients of states with two connected
sites, i.e. states of the form $|\cdots|()|\cdots |$ and the state
$))|\cdots|$ whose coefficients in the eigenvector are denoted by
$\psi_2(x;)$ and $\psi_2(;1,2)$ respectively. The resulting eigenvalue
equations read,
\be
\renewcommand{\arraystretch}{1.2}
\begin{array}{rcl}
\dps \Lambda \psi_2(;) &=& \dps a_- \psi_2(;1) +
\sum_{y=1}^{L-1}\psi_2(y;), \\
\dps \Lambda \psi_2(;1) &=& \dps a_- s_- \psi_2(;1) + \psi_2(1;), \\
\dps \Lambda \psi_2(x;) &=& \dps t \psi_2(x;) + \psi_2(x+1;) +
\psi_2(x-1;),\qquad {\rm for}\;\; 1<x<L-1.
\end{array}
\label{eq:Mn=2a}
\ee
We also find the boundary relations,
\be
\renewcommand{\arraystretch}{1.2}
\begin{array}{rcl}
\dps \Lambda \psi_2(1;) &=& \dps t \psi_2(1;) + \psi_2(2;) +
a_-\psi_2(;1,2),\\ 
\dps \Lambda \psi_2(;1,2) &=& \dps a_- s_- \psi_2(;1,2) +
\psi_2(1;),\\
\dps \Lambda \psi_2(L-1;) &=& \dps t \psi_2(L-1;) + \psi_2(L-2;).
\end{array}
\label{eq:Mn=2d}
\ee
We take the first two lines in (\ref{eq:Mn=2a}) as definitions for
$\psi_2(;)$ and $\psi_2(;1)$, and solve the remaining equations by
making the Ansatz  
\be
\psi_2(x;) = f(x),
\ee
where,
\be
f(x) = A^+ z^x + A^- z^{-x}.
\ee
We then find from (\ref{eq:Mn=2a}) and first line in (\ref{eq:Mn=2d}) that
\be
\renewcommand{\arraystretch}{1.2}
\begin{array}{rcl}
\dps \Lambda &=& \dps t + z+z^{-1},\\
\dps \psi_2(;1,2) &=& \dps a_-^{-1}f(0).
\end{array}
\ee
The other two boundary equations in (\ref{eq:Mn=2d}) imply the
consistency equations, 
\be
\renewcommand{\arraystretch}{1.2}
\begin{array}{rcl}
\dps (\lambda(z)-a_-s_-)f(0) &=& \dps a_-f(1),\\
\dps f(L) &=& \dps 0.
\end{array}
\ee
From these we find equations for the amplitudes
$A^\pm$ and the complex number $z$. In this sector we thus find
that the spectrum is given by 
\be
\Lambda = \lambda(z) = t+z+z^{-1},\qquad z^{2L} = -
\frac{K_+(z)}{K_+(z^{-1})} \frac{A^-}{A^+} 
= \frac{K_+(z)K_-(z)}{K_+(z^{-1})K_-(z^{-1})},
\label{Mn=2BA}
\ee
where,
\be
K_+(z)=1,\qquad K_-(z) = \lambda(z)-a_-(s_-+z).
\ee

\subsection{$n=3$}

From now on we will only look at the equations relating elements of
the eigenvector in the sector $n$ to each other. The equations
involving the elements of sectors $m<n$ can be regarded as definitions
for those elements. The elements in sectors $m>n$ are taken to be zero.

The eigenvalue equation for the sector $n=3$ then reads,
\be
\Lambda \psi_3(x;1) = (a_-s_-+t) \psi_3(x;1) + \psi_3(x+1;1) +
\psi_3(x-1;1)\qquad {\rm for}\;\; 2<x<L,
\label{eq:Mn=3a}
\ee
with boundary relations
\be
\renewcommand{\arraystretch}{1.2}
\begin{array}{rcl}
\dps \Lambda \psi_3(2;1) &=& \dps (a_-s_-+t) \psi_3(2;1) + \psi_3(3;1)
+ \psi_3(1;3),\\ 
\dps \Lambda \psi_3(1;3) &=& \dps t \psi_3(1;3) + \psi_3(2;1) +
a_-\psi_3(;1,2,3),\\ 
\dps \Lambda \psi_3(;1,2,3) &=& \dps a_-s_- \psi_3(;1,2,3) + \psi_3(1;3) +
s_-\psi_3(2;1),\\
\dps \Lambda \psi_3(L-1;1) &=& \dps (a_-s_-+t) \psi_3(L-1;1) +
\psi_3(L-2;1).
\end{array}
\label{eq:Mn=3b}
\ee
Equations (\ref{eq:Mn=3a}) and (\ref{eq:Mn=3b}) can be satisfied by making
the Ansatz,
\be
\psi_3(x;1) = f(x),\qquad f(x) = B^+ z^x + B^- z^{-x}.
\ee
We then find
\be
\renewcommand{\arraystretch}{1.2}
\begin{array}{rcl}
\dps \Lambda &=& \dps a_-s_-+t + z+z^{-1},\\
\dps \psi_3(1;3) &=& \dps f(1),\\
\dps \psi_3(;1,2,3) &=& \dps a_-^{-1} \left(f(0) + a_-s_-
f(1)\right),
\end{array}
\ee
with the consistency equations
\be
\renewcommand{\arraystretch}{1.2}
\begin{array}{rcl}
\dps (\lambda(z)+a_-s_-) f(0) &=& \dps a_-(1-s_-t)f(1),\\
\dps f(L) &=& \dps 0.
\end{array}
\ee
Solving for $B^\pm$ and $z$ we find that in this sector the spectrum
is given by
\be
\Lambda = a_-s_-+\lambda(z)
\ee
where $z$ is a complex number satisfying,
\be
z^{2L} = - \frac{\tilde{K}_+(z)}{\tilde{K}_+(z^{-1})} \frac{B^-}{B^+}
= \frac{\tilde{K}_+(z)\tilde{K}_-(z)}
{\tilde{K}_+(z^{-1})\tilde{K}_-(z^{-1})}
\label{Mn=3BA}
\ee
where
\be
\tilde{K}_+(z)=1,\qquad \tilde{K}_-(z) = \lambda(z)+a_-(s_-+z(s_-t-1)).
%K_-(z)-a_-z((1-s_-)z-s_-t).
\ee

\subsection{general $n$}

From the previous sections we conclude that both sectors $n=2$
and $n=3$ contain only one pseudo particle. There is therefore an
additional quantum number, namely the parity of the number of loop
lines connected to the left boundary, that distinguishes between these
two sectors. As for closed boundaries in Section \ref{se:Cn=4}, we
have checked the sectors containing two pseudo particles (here $n=4$
and $n=5$). In both these sectors, the eigenvalue equation can be  
solved by an Ansatz similar to the one used for closed boundaries, see
(\ref{eq:Cn=2Ansatz}). The Bethe Ansatz equations (\ref{Mn=2BA}) for
the even sector and (\ref{Mn=3BA}) for the odd sector now get
multiplied by the two particle scattering factors $S(z_i,z_j)$, in
much the same way as (\ref{eq:Cn=4BAE}). Furthermore, we have checked
that the Bethe Ansatz equations remain valid in the case when all sites
are connected, i.e when $n=L$.

Because the Hamiltonian $H_L^{\rm M}$ is integrable, the Bethe Ansatz
program can be consistently carried out for arbitrary values of $n$,
and the higher sector equations factorize into those for two
particles. Integrability assures us that we can find consistent
expressions for the more complicated ``merged pseudo particles'' (for
the spectrum we don't need to know these explicitly). 

Before giving the results for general $n$ we first recall the
definitions of $S$ and $\lambda$,
\be
S(z,w) = 1+tw+zw,\qquad \lambda(z) = t+z+z^{-1}.
\ee

\subsubsection{even $n$}

Having derived the equations for $n=2$ and $n=4$, we postulate the
generalization of (\ref{eq:Cn=4BAE}) for arbitrary but even values of
$n$ to the case of mixed boundary conditions,
\be
z_i^{2L} = \frac{K_+(z)K_-(z)}{K_+(z^{-1})K_-(z^{-1})}
\prod_{\stackrel{\scriptstyle j=1}{j \neq i}}^{n/2} 
\frac{S(z_i^{-1},z_j)S(z_j,z_i)}{S(z_j,z_i^{-1})S(z_i,z_j)},
\label{eq:Mn=evenBAE}
\ee
and
\be
K_+(z)=1,\qquad K_-(z) = \lambda(z)-a_-(s_-+z).
\ee
In terms of the solutions of (\ref{eq:Mn=evenBAE}), the eigenvalues of
$H_L^{\rm M}$ are given by, 
\be
\Lambda = \sum_{j=1}^{n/2} \lambda(z_j).
\ee
Using the parametrisation of Section \ref{se:XXZrep} and the
transformation (\ref{eq:z2u}), these equations are equivalent to
the first line of (\ref{eq:BAneven1}) together with (\ref{eq:BAneven2})
in the limit $\beta_+,\alpha_+ \rightarrow \infty$.

\subsubsection{odd $n$}

For $n$ an odd number we postulate the solution from the
investigations of $n=1$, $n=3$ and $n=5$. The eigenvalue for odd $n$
is given by,
\be
\Lambda = a_-s_- + \sum_{j=1}^{(n-1)/2} \lambda(z_j),
\ee
and the numbers $z_i$ satisfy,
\be
z_i^{2L} = \frac{\tilde{K}_+(z)\tilde{K}_-(z)}{\tilde{K}_+(z^{-1})\tilde{K}_-(z^{-1})}
\prod_{\stackrel{\scriptstyle j=1}{j \neq i}}^{n/2} 
\frac{S(z_i^{-1},z_j)S(z_j,z_i)}{S(z_j,z_i^{-1})S(z_i,z_j)},
\label{eq:Mn=oddBAE}
\ee
where
\be
\tilde{K}_+(z)=1,\qquad \tilde{K}_-(z) = \lambda(z)+a_-(s_-+z(s_-t-1)).
\ee
Using the parametrisation of Section \ref{se:XXZrep} and the
transformation (\ref{eq:z2u}), these equations are equivalent to
the first line of (\ref{eq:BAnodd1}) together with (\ref{eq:BAnodd2})
in the limit $\beta_+,\alpha_+ \rightarrow \infty$.

\sectiona{Open boundaries}
\label{se:open}

To diagonalise the Hamiltonian $H_L^{\rm O}$ on the space $V_L^{\rm
O*}$ a similar program can be carried out as for $H_L^{\rm C}$ and
$H_L^{\rm M}$. When the number of connected loops is less than the
system size, i.e. $n<L$, there is no fundamental difference with the
two previous cases. The main difference with the case of mixed
boundaries is that now $K_+(z)$ and $\tilde{K}_+(z)$ become nontrivial
functions. However, when $n=L$ there arises a difficulty which we will
only partially resolve in this paper.

In the following we will first consider the cases when $n<L$ and we
will turn our attention to $n=L$ later. We will denote by
$\psi_{2k+l+m}(x_1,\ldots,x_k;y_1,\ldots,y_l;\tilde{y}_1,\ldots,\tilde{y}_m)$
a coefficient in the eigenvector of a state with $k$ well separated
pseudo particles at positions $x_i$ ($i=1,\ldots,k$), $l$ sites at
positions $y_i$ ($i=1,\ldots,l$) connected to the left boundary, and $m$ sites at
positions $\tilde{y}_i$ ($i=1,\ldots,m$) connected to the right boundary. 
\subsection{$n=0$}

As before, we start with the state with all sites disconnected,
i.e. $\bra{} = |\cdots|$, and find that it is a left eigenstate of
$H_L^{\rm O}$ with eigenvalue $\Lambda=0$.

\subsection{$n=1$}

For $n=1$ we consider eigenvectors in which the coefficient $\psi_1(;;)$ of
the pseudo vacuum is nonzero as well as $\psi_1(;1;)$ of the state
$)|\cdots|$ with one site connected to the left boundary and
$\psi_1(;;L)$ of the state $|\cdots|($ with one site connected to the
right boundary. The eigenvalue equations are 
\be
\renewcommand{\arraystretch}{1.2}
\begin{array}{rcl}
\dps \Lambda \psi_1(;;) &=& \dps a_- \psi_1(;1;)+a_+\psi_1(;;L),\\
\dps \Lambda \psi_1(;1;) &=& \dps a_- s_- \psi_1(;1;),\\
\dps \Lambda \psi_1(;;L) &=& \dps a_+ s_+ \psi_1(;;L).
\end{array}
\label{eq:On=1a}
\ee
As before we regard the first line in (\ref{eq:On=1a}) as the
definition of $\psi_1(;;)$. In the generic case of $a_-s_- \neq
a_+s_+$, the remaining equations are solved by either
\be
\Lambda=a_- s_-,\qquad \psi_1(;;L)=0,
\label{eq:On=1b}
\ee
or
\be
\Lambda=a_+ s_+,\qquad \psi_1(1;;)=0.
\label{eq:On=1c}
\ee

\subsection{$n=2$}

In this sector all states with two connected sites, i.e. states of the
form $|\cdots|()|\cdots |$, $))|\cdots|$ and $|\cdots|(($, may have
nonzero coefficients in the eigenvector. These coefficients are
denoted by $\psi_2(x;;)$, $\psi_2(;1,2;)$ and $\psi_2(;;L-1,L)$
respectively. The resulting eigenvalue equations read,
\be
\renewcommand{\arraystretch}{1.2}
\begin{array}{rcl}
\dps \Lambda \psi_2(;;) &=& \dps a_- \psi_2(;1;) + a_+ \psi_2(;;L) +
\sum_{y=1}^{L-1}\psi_2(y;;), \\
\dps \Lambda \psi_2(;1;) &=& \dps a_- s_- \psi_2(;1;) + a_+
\psi_2(;1;L) + \psi_2(1;;), \\
\dps \Lambda \psi_2(;;L) &=& \dps a_+ s_+ \psi_2(;;L) + a_-
\psi_2(;1;L) + \psi_2(L-1;;), \\
\dps \Lambda \psi_2(x;;) &=& \dps t \psi_2(x;;) + \psi_2(x+1;;) +
\psi_2(x-1;;),\qquad {\rm for}\;\; 1<x<L-1.
\end{array}
\label{eq:On=2a}
\ee
In addition, we also find the boundary relations,
\be
\renewcommand{\arraystretch}{1.2}
\begin{array}{rcl}
\dps \Lambda \psi_2(1;;) &=& \dps t \psi_2(1;;) + \psi_2(2;;) +
a_-\psi_2(;1,2;),\\ 
\dps \Lambda \psi_2(L-1;;) &=& \dps t \psi_2(L-1;;) + \psi_2(L-2;;) +
a_+ \psi_2(;;L-1,L),\\
\dps \Lambda \psi_2(;1,2;) &=& \dps a_- s_- \psi_2(;1,2;) +
\psi_2(1;;),\\
\dps \Lambda \psi_2(;;L-1,L) &=& \dps a_+ s_+ \psi_2(;;L-1,L) +
\psi_2(L-1;;),\\
\dps \Lambda \psi_2(;1;L) &=& \dps (a_-s_-+a_+ s_+)\psi_2(;1;L).
\end{array}
\label{eq:On=2d}
\ee
As before, we take the first three lines in (\ref{eq:On=2a}) as definitions for
$\psi_2(;;)$, $\psi_2(;1;)$ and $\psi_2(;;L)$, and solve the remaining
equations by making the Ansatz  
\be
\psi_2(x;;) = f(x),
\ee
where,
\be
f(x) = A^+ z^x + A^- z^{-x}.
\ee
We then find from the last line in (\ref{eq:On=2a}) and the first two
lines in (\ref{eq:On=2d}) that 
\be
\renewcommand{\arraystretch}{1.2}
\begin{array}{rcl}
\dps \Lambda &=& \dps t + z+z^{-1},\\
\dps \psi_2(;1,2;) &=& \dps a_-^{-1}f(0),\\
\dps \psi_2(;;L-1,L) &=& \dps a_+^{-1}f(L).
\end{array}
\label{eq:On=2Lam}
\ee
The third and fourth boundary equations in (\ref{eq:On=2d}) imply the
consistency equations, 
\be
\renewcommand{\arraystretch}{1.2}
\begin{array}{rcl}
\dps (\lambda(z)-a_-s_-)f(0) &=& \dps a_-f(1),\\
\dps (\lambda(z)-a_+s_+)f(L) &=& \dps a_+f(L-1).
\end{array}
\ee
In the case of mixed boundaries we saw that we needed to distinguish
the case with an odd number from that with an even number of loops
connected to the left boundary. The same holds true here, as can be
seen by the last equation of (\ref{eq:On=2d}), which implies that
$\psi_2(;1;L)=0$ unless $\Lambda = a_-s_-+a_+s_+$, which will
generically not be the case with $\Lambda$ given by
(\ref{eq:On=2Lam}). The complete spectrum in this sector is therefore
given by 
\be
\Lambda = a_-s_-+a_+s_+,
\ee
in which case $A^-=A^+=0$ and $\psi_2(;1;L)\neq 0$, or
\be
\Lambda = \lambda(z) = t+z+z^{-1},
\label{On=2Lam2}
\ee
in which case $\psi_2(;1;L)= 0$. The complex number $z$ in
(\ref{On=2Lam2}) satisfies the Bethe Ansatz equation
\be
z^{2L} = - \frac{K_+(z)}{K_+(z^{-1})} \frac{A^-}{A^+} 
= \frac{K_+(z)K_-(z)}{K_+(z^{-1})K_-(z^{-1})},
\ee
where,
\be
K_\pm(z)=\lambda(z)-a_\pm(s_\pm+z).
\ee

\subsection{$n=3$}

In analogy with the case of mixed boundaries, we do not write
anymore explicitly the equations relating eigenvector elements of
sectors smaller than $n$ to those of $n$, but only the equations
relating the latter to each other. The eigenvalue equation then reads,
\be
\renewcommand{\arraystretch}{1.2}
\begin{array}{rcl}
\Lambda \psi_3(x;1;) &=& (a_-s_-+t) \psi_3(x;1;) + \psi_3(x+1;1;) +
\psi_3(x-1;1;),\\ && {\rm for}\;\; 2<x<L-1,\\
\Lambda \psi_3(x;;L) &=& (a_+s_++t) \psi_3(x;;L) + \psi_3(x+1;;L) +
\psi_3(x-1;;L),\\ 
&& {\rm for}\;\; 1<x<L-2.
\end{array}
\label{eq:On=3a}
\ee
with the additional relations when an odd number of loops is
connected to the left boundary,
\be
\renewcommand{\arraystretch}{1.2}
\begin{array}{rcl}
\dps \Lambda \psi_3(2;1;) &=& \dps (a_-s_-+t) \psi_3(2;1;) + \psi_3(3;1;)
+ \psi_3(1;3;),\\ 
\dps \Lambda \psi_3(1;3;) &=& \dps t \psi_3(1;3;) + \psi_3(2;1;) +
a_-\psi_3(;1,2,3;),\\ 
\dps \Lambda \psi_3(;1,2,3;) &=& \dps a_-s_- \psi_3(;1,2,3;) + \psi_3(1;3;) +
s_-\psi_3(2;1;),\\
\dps \Lambda \psi_3(L-1;1;) &=& (a_-s_-+t) \psi_3(L-1;1;) +
\psi_3(L-2;1;) \nonumber \\
&&{} + a_+ \psi_3(;1;L-1,L),\\
\dps \Lambda \psi_3(;1;L-1,L) &=&  (a_-s_-+a_+s_+) \psi_3(;1;L-1,L) + \psi_3(L-1;1;).
\end{array}
\label{eq:On=3b}
\ee
and similar equations when an even number of loops is connected to the
left boundary.
These equations can be satisfied by making the Ansatz,
\be
\renewcommand{\arraystretch}{1.2}
\begin{array}{rcl}
\psi_3(x;1;) = f(x),\qquad f(x) = C^+z^x + C^- z^{-x},\\
\psi_3(x;;L) = g(x),\qquad g(x) = D^+z^x + D^- z^{-x}.
\end{array}
\ee
Since the equations do not mix states with an even and odd number of
loops connected to the left boundary, we find they can be satisfied if
either $D^\pm=0$ or $C^\pm=0$. In the first case we have,
\be
\renewcommand{\arraystretch}{1.2}
\begin{array}{rcl}
\dps \Lambda &=& \dps a_-s_-+t + z+z^{-1},\\
\dps \psi_3(1;3;) &=& \dps f(1),\\
\dps \psi_3(;1,2,3;) &=& \dps a_-^{-1} \left(f(0) + a_-s_-
f(1))\right),\\
\dps \psi_3(;1;L-1,L) &=& a_+^{-1} f(L),
\end{array}
\label{eq:Csol}
\ee
with the consistency equations
\be
\renewcommand{\arraystretch}{1.2}
\begin{array}{rcl}
\dps (\lambda(z)+a_-s_-) f(0) &=& \dps a_-(1-s_-t)f(1),\\
\dps (\lambda(z)-a_+s_+)f(L) &=& \dps a_+ f(L-1).
\end{array}
\ee
Solving for $C^\pm$ and $z$ we find that in this sector one part of
the spectrum is given by the first line of (\ref{eq:Csol})
where $z$ is a complex number satisfying,
\be
z^{2L} = - \frac{K_+(z)}{K_+(z^{-1})} \frac{C^-}{C^+}
= \frac{K_+(z)\tilde{K}_-(z)}
{K_+(z^{-1})\tilde{K}_-(z^{-1})}
\label{On=3BA1}
\ee
where
\be
K_+(z)=\lambda(z) - a_+(s_++z),\qquad \tilde{K}_-(z) = \lambda(z)+a_-(s_-+z(s_-t-1)).
\ee

Solving for $D^\pm$ and $z$ in the case where $C^\pm=0$, we find that
the other part of the spectrum is given by
\be
\Lambda = a_+s_++\lambda(z)
\ee
where $z$ is a complex number now satisfying,
\be
z^{2L} = - \frac{\tilde{K}_+(z)}{\tilde{K}_+(z^{-1})} \frac{D^-}{D^+}
= \frac{\tilde{K}_+(z)K_-(z)}
{\tilde{K}_+(z^{-1})K_-(z^{-1})}
\label{On=3BA2}
\ee
with
\be
\tilde{K}_+(z)=\lambda(z)+a_+(s_++z(s_+t-1)),\qquad K_-(z) = \lambda(z)-a_-(s_-+z).
%K_-(z)-a_-z((1-s_-)z-s_-t).
\ee

\subsection{general $n$}
\label{se:Ogenn}

Sectors higher than $n=3$ introduce multi particle scattering. The
spectrum in these sectors can be calculated using a similar Ansatz as
for mixed boundaries. While the equations become more cumbersome, the
final answer is the obvious generalisation of the previous sectors if
we keep the results for mixed boundaries in mind. The results below
are valid for $n<L$. 

Before giving the final result, we first recall the definitions
\be
\renewcommand{\arraystretch}{1.2}
\begin{array}{rcl}
\lambda(z) &=& t+z+z^{-1},\\
S(z,w) &=& 1+tw+zw,\\
K_\pm(z) &=& \lambda(z)-a_\pm(s_\pm+z),\\
\tilde{K}_\pm(z) &=& \lambda(z)+a_\pm(s_\pm+z(s_\pm t-1)).
\end{array}
\ee

\subsubsection{even $n$}

The spectrum in these sectors is described by the following two cases. 
\begin{itemize}
\item Odd number of loops to both boundaries.
\be
\Lambda = a_-s_- + a_+s_+ + \sum_{j=1}^{n/2-1} \lambda(z_j),
\ee
where the numbers $z_i$ satisfy,
\be
z_i^{2L} = \frac{\tilde{K}_+(z)\tilde{K}_-(z)}{\tilde{K}_+(z^{-1})\tilde{K}_-(z^{-1})}
\prod_{\stackrel{\scriptstyle j=1}{j \neq i}}^{n/2-1} 
\frac{S(z_i^{-1},z_j)S(z_j,z_i)}{S(z_j,z_i^{-1})S(z_i,z_j)},
\label{eq:On=evenBAE1}
\ee

\item Even number of loops to both boundaries.
\be
\Lambda = \sum_{j=1}^{n/2} \lambda(z_j),
\ee
where the numbers $z_i$ satisfy,
\be
z_i^{2L} = \frac{K_+(z)K_-(z)}{K_+(z^{-1})K_-(z^{-1})}
\prod_{\stackrel{\scriptstyle j=1}{j \neq i}}^{n/2} 
\frac{S(z_i^{-1},z_j)S(z_j,z_i)}{S(z_j,z_i^{-1})S(z_i,z_j)}.
\label{eq:On=evenBAE2}
\ee
\end{itemize}
Using the parametrisation of Section \ref{se:XXZrep} and the
transformation (\ref{eq:z2u}), these equations are equivalent to
(\ref{eq:BAneven1})-(\ref{eq:BAneven3}).

\subsubsection{odd $n$}

For these sectors the spectrum also falls into two classes. 
\begin{itemize}
\item Odd number of loops to the left and even number of loops
  connected to the right boundary.
\be
\Lambda = a_-s_- + \sum_{j=1}^{(n-1)/2} \lambda(z_j),
\ee
where the numbers $z_i$ satisfy,
\be
z_i^{2L} = \frac{K_+(z)\tilde{K}_-(z)}{K_+(z^{-1})\tilde{K}_-(z^{-1})}
\prod_{\stackrel{\scriptstyle j=1}{j \neq i}}^{(n-1)/2} 
\frac{S(z_i^{-1},z_j)S(z_j,z_i)}{S(z_j,z_i^{-1})S(z_i,z_j)},
\label{eq:On=oddBAE1}
\ee
\item Even number of loops to both boundaries.
\be
\Lambda = a_+s_+ + \sum_{j=1}^{(n-1)/2} \lambda(z_j),
\ee
where the numbers $z_i$ satisfy,
\be
z_i^{2L} = \frac{\tilde{K}_+(z)K_-(z)}{\tilde{K}_+(z^{-1})K_-(z^{-1})}
\prod_{\stackrel{\scriptstyle j=1}{j \neq i}}^{(n-1)/2} 
\frac{S(z_i^{-1},z_j)S(z_j,z_i)}{S(z_j,z_i^{-1})S(z_i,z_j)}.
\label{eq:On=oddBAE2}
\ee
\end{itemize}
Using the parametrisation of Section \ref{se:XXZrep} and the
transformation (\ref{eq:z2u}), these equations are equivalent to
(\ref{eq:BAnodd1})-(\ref{eq:BAnodd3}).

\subsection{$n=L=1$}
\label{se:n=L=1}

We now consider the cases where $n=L$. These sectors are quite
different from the previous cases since the nonlocal additional
relations (\ref{eq:TLb}) involving the parameter $b$ come into play,
see also section \ref{se:looprep} and section \ref{se:BAintro}. In this sense
they are similar to the $n=L$  sectors in the periodic TL algebra, see
e.g. \cite{MartinS93}. 

Recall that for $n=L$ we only consider the subspace generated by
$I_L\ket{}$, see section \ref{se:looprep}. The eigenvalue equations 
in the $n=L=1$ sector are 
\be
\renewcommand{\arraystretch}{1.2}
\begin{array}{rcl}
\dps \Lambda \psi_1(;1;) &=& \dps a_- s_- \psi_1(;1;) + a_+ \psi_1(;;1),\\
\dps \Lambda \psi_1(;;1) &=& \dps a_+ s_+ \psi_1(;;1) + a_-b \psi_1(;1;).
\end{array}
\label{eq:On=L=1a}
\ee
These equations are easily solved, and the spectrum is given by the
solutions of the characteristic equation
\be
(\Lambda-a_- s_-)(\Lambda-a_+s_+)=a_-a_+ b.
\ee
Hence we see that if $b=0$ we have the same solutions (\ref{eq:On=1b})
and (\ref{eq:On=1c}) as in the case $n=1<L$. We also note that
$b=s_-s_+$ is another special value where the solutions do not contain
a radical (see also the remarks at the end of Section \ref{se:result}).  

\subsection{$n=L=2$}

In this sector, the resulting eigenvalue equations are,
\be
\renewcommand{\arraystretch}{1.2}
\begin{array}{rcl}
\dps \Lambda \psi_2(1;;) &=& \dps t \psi_2(1;;) + a_-\psi_2(;1,2;) +
a_+\psi_2(;;1,2),\\ 
\dps \Lambda \psi_2(;1,2;) &=& \dps a_- s_- \psi_2(;1,2;) +
\psi_2(1;;) +a_+ \psi_2(;1;2),\\
\dps \Lambda \psi_2(;;1,2) &=& \dps a_+ s_+ \psi_2(;;1,2) +
\psi_2(1;;) +a_- \psi_2(;1;2),\\
\dps \Lambda \psi_2(;1;2) &=& \dps (a_-s_-+a_+ s_+)\psi_2(;1;2) + b\psi_2(1;;).
\end{array}
\label{eq:On=L=2d}
\ee
For $b=0$, these equations can be solved in the same way as for
$n=2<L$ by making the Ansatz,
\be
\psi_2(x;;) = f(x),
\ee
where,
\be
f(x) = A^+ z^x + A^- z^{-x}.
\ee
We then find that the complete spectrum in this sector is given by
\be
\Lambda = a_-s_-+a_+s_+,
\label{On=L=2Lam1}
\ee
in which case one can take $\psi_2(;1;2)\neq 0$, or 
\be
\Lambda = \lambda(z) = t+z+z^{-1} \neq a_-s_-+a_+s_+,
\label{On=L=2Lam2}
\ee
in which case $\psi_2(;1;2)= 0$. The complex number $z$ in
(\ref{On=L=2Lam2}) satisfies the Bethe Ansatz equation
\be
z^{2L} = - \frac{K_+(z)}{K_+(z^{-1})} \frac{A^-}{A^+} 
= \frac{K_+(z)K_-(z)}{K_+(z^{-1})K_-(z^{-1})},
\ee
where,
\be
K_\pm(z)=\lambda(z)-a_\pm(s_\pm+z).
\ee

The eigenvalue is a zero of a quartic polynomial. For $b=0$ this
polynomial factorizes into a cubic and a linear polynomial. The zero
of the linear polynomial gives the solution (\ref{On=L=2Lam1}). If
$b\neq 0$, the eigenvalue is a solution of a quartic equation that
does not factorize in general. We have not found a systematic way of
solving the case $b\neq 0$ and $n=L$.

\subsection{$n=L=3$}

In this sector there are eight equations resulting from the eigenvalue equation,
\be
\renewcommand{\arraystretch}{1.2}
\begin{array}{rcl}
\dps \Lambda \psi_3(2;1;) &=& \dps (a_-s_-+t) \psi_3(2;1;) +
\psi_3(1;3;) + a_+\psi_3(;1;2,3),\\ 
\dps \Lambda \psi_3(1;3;) &=& \dps t \psi_3(1;3;) + \psi_3(2;1;) +
a_-\psi_3(;1,2,3;) + a_+\psi_3(1;;3),\\ 
\dps \Lambda \psi_3(;1;2,3) &=&  (a_-s_-+a_+s_+) \psi_3(;1;2,3) +
\psi_3(2;1;) + \psi_3(1;;3),\\
\dps \Lambda \psi_3(;1,2,3;) &=& \dps a_-s_- \psi_3(;1,2,3;) + \psi_3(1;3;) +
s_-\psi_3(2;1;) +a_+\psi_3(;1,2;3),
\end{array}
\label{eq:On=L=3a}
\ee
and
\be
\renewcommand{\arraystretch}{1.2}
\begin{array}{rcl}
\dps \Lambda \psi_3(1;;3) &=& (a_+s_++t) \psi_3(1;;3) +
\psi_3(2;;1) + a_-\psi_3(;1,2;3) \nonumber \\
\dps \Lambda \psi_3(2;;1) &=& \dps t \psi_3(2;;1) + \psi_3(1;;3) +
a_-b \psi_3(2;1;) + a_+ \psi_3(;;1,2,3), \nonumber\\
\dps \Lambda \psi_3(;1,2;3) &=& (a_-s_- + a_+s_+)  \psi_3(;1,2;3) +
\psi_3(1;;3) + b \psi_3(2;1;),\nonumber\\
\dps \Lambda \psi_3(;;1,2,3) &=& a_+s_+ \psi_3(;;1,2,3) + \psi_3(2;;1)
+ s_+ \psi_3(1;;3) + a_-b \psi_3(;1;2,3).
\end{array}
\label{eq:On=L=3b}
\ee
We will try to solve them in a similar way as for $n=3<L$. First write,
\be
\renewcommand{\arraystretch}{1.2}
\begin{array}{rclrclrcl}
\dps \psi_3(2;1;) &=& \dps f(2),& \dps \psi_3(1;3;) &=& \dps
f(1),\qquad & f(x) &=& C^+z^x + C^- z^{-x},\\ 
\dps \psi_3(1;;3)&=& \dps g(1),& \psi_3(2;;1)&=& \dps g(2), &\dps g(x) &=&
\dps D^+z^x + D^- z^{-x}.
\end{array}
\ee
If $b=0$, the equations (\ref{eq:On=L=3b}) only contain coefficients
of states with an even number of loops connected to the left
boundary. Hence (\ref{eq:On=L=3b}) will be trivially satisfied if we
set these coefficients all to zero, i.e. $D^+=D^-=0$. The equations
(\ref{eq:On=L=3a}) can then be solved in a similar way as for the case
$n=3<L$, and we find,
\be
\Lambda = a_-s_-+\lambda(z),
\ee
where $z$ is a complex number satisfying,
\be
z^{2L} = - \frac{K_+(z)}{K_+(z^{-1})} \frac{C^-}{C^+}
= \frac{K_+(z)\tilde{K}_-(z)}
{K_+(z^{-1})\tilde{K}_-(z^{-1})},
\label{On=L=3BA1}
\ee
and
\be
K_+(z)=\lambda(z) - a_+(s_++z),\qquad \tilde{K}_-(z) = \lambda(z)+a_-(s_-+z(s_-t-1)).
\ee

Another solution for $b=0$ is obtained in the following way. We find
a non-trivial solution of the eigenvalue problem
(\ref{eq:On=L=3b}), which exists provided one satisfies,
\be
\Lambda = a_+s_++\lambda(z),
\ee
where $z$ is a complex number satisfying,
\be
z^{2L} = - \frac{\tilde{K}_+(z)}{\tilde{K}_+(z^{-1})} \frac{D^-}{D^+}
= \frac{\tilde{K}_+(z)K_-(z)}
{\tilde{K}_+(z^{-1})K_-(z^{-1})},
\label{On=L=3BA2}
\ee
and
\be
\tilde{K}_+(z)=\lambda(z)+a_+(s_++z(s_+t-1)),\qquad K_-(z) = \lambda(z)-a_-(s_-+z).
\ee
Given such a solution, the four equations (\ref{eq:On=L=3a}) then
contain four unknowns, the four coefficients with an even number of
loops connected to the left boundary. Apart from accidental
degeneracies, (\ref{eq:On=L=3a}) thus has a solution, that we do not
need to know explicitly here. Hence for $b=0$, the solution for
$n=L=3$ is the same as that for $n=3<L$. 

\subsection{general $n=L$}

We hope to have convinced the reader that for $b=0$, the Bethe Ansatz
solution for the spectrum for the case $n<L$ in Section \ref{se:Ogenn}
remains valid also for $n=L$. 

\section{Conclusion}

We have derived equations for the spectrum of the Temperley-Lieb loop
model with closed, mixed and open boundaries using a coordinate Bethe
Ansatz. The spectra of the quantum XXZ spin chain with diagonal, one
non-diagonal and two non-diagonal open boundaries respectively are
contained in those of the loop model. In the case of two non-diagonal
boundaries, we thus derive a recent numerically conjectured result,
and obtain the spectrum of the spin chain in a part of the parameter
space where it was previously not known. 

\section{Acknowledgement}
It is a pleasure to thank Vladimir Rittenberg for many useful
discussions, and Andrew Rechnitzer for his help in obtaining
the results of the appendix. JdG was financially supported
by the Australian Research Council. PP was partially supported by the
Heisenberg-Landau grant and by the RFBR grant No.03-01-00781.  

\appendix

\section{Dimension of representation spaces}

We use a functional equation method \cite{PrellB95} to find the
generating functions of the various dimensions of the representation
spaces. Let us first consider all sequences of well-nested parentheses
that have the same number of opening and closing parentheses. Such a
sequence may be the empty sequence, or it has the form
$\bullet(\bullet)$ where $\bullet$ represents any well-nested sequence
of parentheses, including the empty one. The generating function for
such sequences hence satisfies the functional equation 
\be
D(z) = 1 + z^2 D(z)^2,
\label{eq:Dyck}
\ee
where the $1$ coresponds to the empty sequence, $z^2$ to the opening
and closing parenthesis and $D(z)^2$ to the two well-nested
sequences. Equation (\ref{eq:Dyck}) is easily solved with the initial
condition $D(0)=1$. The solution is the Catalan generating function
\be
D(z) = \frac{2}{1+\sqrt{1-4z^2}}.
\ee

Let us now consider all sequences of well-nested parenthesis with an
arbitrary number of vertical bars ``$|$'' in between them. Each such
sequence with $n$ vertical bars is of the form
$$\bullet|\bullet|\cdots\bullet|\bullet.$$ 
The generating function for these sequences is therefore given by  
\bea
G_{\rm C}(z) &=& D(z) + D(z)zD(z) + D(z)zD(z)zD(z) = \ldots \\
&=& D(z) \sum_{n=0}^\infty (zD(z))^n = \frac{D(z)}{1-zD(z)}.
\eea

The sequences corresponding to states in $V_L^{\rm M}$ are of the form
$$\bullet)\bullet)\cdots\bullet)\bullet|\bullet|\cdots\bullet|\bullet,$$
and hence the generating function is given by
\be
G_{\rm M}(z) = D(z) \left(\sum_{n=0}^\infty (zD(z))^n\right)^2 = \frac{D(z)}{(1-zD(z))^2} =
\frac{1}{1-2z}.
\label{eq:GM(z)}
\ee

Finally, the sequences corresponding to states in $V_L^{\rm O}$ are of the form
$$\bullet)\bullet)\cdots\bullet)\bullet|\bullet|\cdots\bullet|\bullet(\bullet\cdots
(\bullet(\bullet,$$ 
and hence the generating function is given by
\be
G_{\rm O}(z) = D(z) \left(\sum_{n=0}^\infty (zD(z))^n\right)^3 = \frac{D(z)}{(1-zD(z))^3}.
\ee
The generating functions for the subspaces generated by $I_L\ket{}$ and
$J_L\ket{}$, see section \ref{se:looprep}, are equal to $G_{\rm
  M}(z)$, which follows from replacing ``$|$'' by ``$($'' in the
sequences above (\ref{eq:GM(z)}).

\end{document}